\def\3{{\ss}}
\newcommand{\beq}{\begin{equation}}
\newcommand{\eeq}{\end{equation}}
\newcommand{\beqa}{\begin{eqnarray}}
\newcommand{\eeqa}{\end{eqnarray}}

\documentclass[epj]{svjour}
\usepackage{latexsym}
\usepackage{graphicx}
\begin{document}
\title{Updated dispersion-theoretical analysis of the nucleon
electromagnetic form factors\thanks{Work supported in part by the DFG 
under contract HA 3203/2-1.}
}
\author{H.-W. Hammer\inst{1}$^,$\thanks{\emph{Present address:}
Institute for Nuclear Theory, University of Washington, Seattle, WA 98195, 
USA}$^,$\thanks{Email: hammer@phys.washington.edu} 
\and Ulf--G.~Mei\3ner\inst{1,2}$^,$\thanks{Email: meissner@itkp.uni-bonn.de} 
}                     
%
%
\institute{Universit\"at Bonn, Helmholtz-Institut f{\"u}r
Strahlen- und Kernhysik (Theorie), D-53115 Bonn, Germany \and 
Forschungszentrum J\"ulich, Institut f\"ur Kernphysik 
(Theorie), D-52425 J\"ulich, Germany}
\date{Received: date / Revised version: date}
%
\abstract{In the light of the new data on the various neutron and proton
electromagnetic form factors taken in recent years, 
we update the dispersion-theoretical 
analysis of the nucleon electromagnetic form factors from the mid-nineties.
The parametrization of the spectral functions includes constraints
from unitarity, perturbative QCD, and recent measurements of the neutron
charge radius. We obtain a good description of most modern form factor
data, with the exception of the Jefferson Lab data on $G_E^p/G_M^p$ in
the four-momentum transfer range $Q^2=3...6$ GeV$^2$.
For the magnetic radii of the proton 
and the neutron we find $r_M^p = 0.857$ fm and $r_M^n = 0.879$ fm, which is 
consistent with the recent determinations using continued fraction
expansions.
} 
\titlerunning{Updated dispersion-theoretical analysis}
\authorrunning{H.-W. Hammer, Ulf-G. Mei{\ss}ner}
\maketitle
\section{Introduction}
\label{intro}

The electromagnetic form factors of the nucleon encode information about the
structure of the nucleon over a wide range of scales. 
Depending on the kinematical conditions, they can be determined most 
precisely from elastic electron scattering using Rosenbluth-separation
or in double polarization experiments. 
Only in the last decade has it become technically feasible to perform the 
double polarization experiments. Thus the data base has been 
considerably enlarged since our last dispersion analysis was performed in
1996 \cite{MMD,HMD}. This holds in particular for the neutron form factors, 
which can only be measured indirectly using deuterium or $^3$He targets. 
For recent reviews, see e.g. Refs.~\cite{Gao,Schmieden}. At low momentum 
transfer, one obtains information about the various nucleon rms-radii, 
which are not only of interest by themselves but e.g. the proton charge 
radius $r_E^p$ has also to be 
known to a good accuracy to perform precision tests of QED in Lamb-shift 
measurements, or conversely, one can use such experiments to pin down 
$r_E^p$. For some time, low-energy  electron scattering and Lamb shift 
determinations led to different values for  $r_E^p$, but this discrepancy 
has been resolved, see Refs.~\cite{Rosenfelder,Sick:2003gm}
and references therein.
One can also extract vector meson--nucleon coupling constants and eventually 
study the transition from the non-perturbative regime of QCD to the 
perturbative one. Since the data span the range of four-momentum transfers from
$Q^2 \simeq 0$ to $Q^2 \simeq 31\,$GeV$^2$,  
the only model-independent method to analyze 
these data is dispersion theory. The most recent 
dispersion-theoretical analysis of the nucleon electromagnetic form factors
dates back to almost a decade ago, including fits to space-like \cite{MMD}
as well as time-like data \cite{HMD}. For some recent vector-meson
pole-model approaches to the nucleon electromagnetic form factors,
see Refs.~\cite{Lomo,Dubni}.
In view of the new data and a new
data base collected and discussed in Ref.~\cite{FW}, it seems timely to update
the work of Refs.~\cite{MMD,HMD} using the data basis of \cite{FW}. 
It is also interesting to investigate whether
such a general scheme can lead to the pronounced structure around 
$Q^2 = 0.2\,$GeV$^2$ found for all four form factors in the 
phenomenological analysis of Ref.~\cite{FW}.

\section{Formalism}
\label{sec:form}
The electromagnetic structure of the nucleon is parameterized 
by the Dirac ($F_1$) and Pauli ($F_2$) form factors for the
proton and the neutron 
\beq
F_i^{p/n} (Q^2) =  F_i^S (Q^2) \pm \,  F_i^V (Q^2)~,\quad i=1,2,
\eeq
with $Q^2$ the four--momentum transfer of the virtual photon
(the photon virtuality). Our conventions are such that
$Q^2 >0$ for space-like momentum transfer.
We have expressed the nucleon form factors in
the isospin basis ($S=$ isoscalar, $V=$ isovector) which
is most appropriate for the dispersive analysis.
The experimental data are usually given for the 
Sachs form factors, which are linear combinations of $F_1$ and 
$F_2$:
\beqa
G_E^I (Q^2) &=& F_1^I (Q^2) - \frac{Q^2}{4m^2} F_2^I (Q^2)~,\\
G_M^I (Q^2) &=& F_1^I (Q^2) + F_2^I (Q^2)~, \quad I = S, V.
\eeqa

The analysis of the  nucleon 
electromagnetic form factors proceeds most directly through the spectral 
representation given 
by\footnote{Note that we work with unsubtracted dispersion relations.
Since the normalizations of the various form factors are known, one could also
work with once--subtracted dispersion relations.}
\beq\label{spec}
F_i^I (Q^2) = \frac{1}{\pi}\, \int_{{(\mu_0^I)}^2}^\infty   
\frac{\sigma_i^I (\mu^2) \, d\mu^2}{\mu^2 + Q^2}~, \quad i =1,2,
\ I = S, V,
\eeq
in terms of the real spectral functions 
$\sigma_i^I (\mu^2) = {\rm Im}\,F_i^I (\mu^2)$,
and the corresponding thresholds are given by 
$\mu_0^S = 3M_\pi$, $\mu_0^V = 2M_\pi$.
The spectral functions encode the pertinent physics of the nucleon form
factors. In the isovector channel, the spectral function is build up
by the two--pion continuum
(including the $\rho$--resonance) as given by unitarity 
\cite{Hohler:1974ht} plus a series
of poles, whose masses and residues are fit parameters. In the isoscalar
channel, we only have poles, where the lowest two are given by the
$\omega$ and the $\phi$ mesons, respectively. We restrict the
number of poles in both channels by the stability condition 
of Ref.~\cite{SabbaStefanescu:1978sy} which requires
to use the minimum number of vector meson poles necessary to fit the 
data. Note that this condition  was also used in the earlier fits
\cite{Hohler:1974ht,MMD,HMD}.
Enforcing the correct normalization conditions for all
form factors, the experimental value of the neutron charge radius and the
superconvergence relations from the perturbative
QCD behaviour of the form factors, e.g.
$F_i (Q^2) \sim 1/(Q^2)^{i+1}$ as $Q^2$ tends to infinity,  
reduces the number of free fit parameters considerably.
For more details on the form of the spectral functions, the
stability condition, and the way the various constraints are included, 
see Ref.~\cite{MMD}. 

It is possible to take all pole masses from physical particles
(except for one which is determined by the constraints discussed above). 
As a consequence, we have 2 (3) free parameters in the isovector
(isoscalar) channel if we restrict the number of poles to 3 (4).
Furthermore, we have one more free parameter that characterizes the 
onset of the leading logarithms from perturbative QCD.  
A difference to the earlier fits \cite{MMD,HMD} is that
the constraint from the neutron charge radius has somewhat changed. While
in  \cite{MMD,HMD} an electron-neutron scattering length of 
$b_{ne} = (-1.308 \pm 0.05) \cdot 10^{-3}$~fm was used, 
the reevaluation of the
data lead to $b_{ne} = (-1.33 \pm 0.027 \pm 0.03) \cdot 10^{-3}$~fm 
for scattering off $^{208}$Pb 
and $b_{ne} = (-1.44 \pm 0.033 \pm 0.06) \cdot 10^{-3}$~fm for $^{209}$Bi 
\cite{Kop}. We have performed fits with both values but only show results 
for the Pb value which seems to be favored by the $G_E^n$ data.

\section{Results}
\label{sec:res}

Before discussing the results, we must specify the data to which we fit. 
We use the data basis collected and specified by Friedrich and Walcher
in Ref.~\cite{FW}. It consists of a total of 190 data points, 
about a quarter of which were not available in 1996 (mainly for the neutron).
Furthermore, the data basis has also been pruned for inconsistent
data and therefore does not make use of some of the data that were 
included in the fits of \cite{MMD,HMD}. 
For a detailed list of the data taken into consideration, we refer 
the reader to Ref.~\cite{FW}. In particular, the Jefferson Lab data
for $G_E^p /G_M^p$ \cite{Jones:1999rz,Gayou:2001qd} are treated as
data for $G_E^p$ in \cite{FW} since $G_M^p$ is supposed to be known 
better for these virtualities.

Fitting these data with the readjusted constraints leads to the following 
parameters. The iso\-sca\-lar mass\-es are $M_\omega = 0.782$ GeV, $M_\phi =
1.019$ GeV, $M_{S'} = 1.65$ GeV, and $M_{S''} = 1.68$ GeV. Note that in
contrast to the earlier fits, we can not work with 3 isoscalar poles only.
The isovector masses are $M_{\rho '} = 1.05$ GeV, $M_{\rho ''} = 1.465$ GeV,
and $M_{\rho '''} = 1.70$ GeV. Note that except from $M_{\rho '}$, whose
value is fixed from the various constraints, all these 
masses correspond to physical particles listed in the PDG tables. The
corresponding residua for the isoscalar poles are:
$a_1^{\omega} = 0.767$, $a_2^{\omega} = 0.318$,
$a_1^{\phi} = -0.832$, $a_2^{\phi} = -0.250$,
$a_1^{S'} = 2.09$, $a_2^{S'} = 3.97$,
$a_1^{S''} = -2.04$, and $a_2^{S''} = -3.76$,
where the subscript 1 (2) refers to the vector (tensor) coupling 
of the corresponding vector meson to the nucleon.
Similarly, we have for the three isovector poles:
$a_1^{\rho'} = -0.154$, $a_2^{\rho'} = -0.306$,
$a_1^{\rho''} = 1.15$, $a_2^{\rho''} = -4.42$,
$a_1^{\rho'''} = -1.32$, and $a_2^{\rho'''} = 3.18$.
The QCD parameters (for definitions, see \cite{MMD}) are $\gamma = 2.148$,
$\Lambda^2 = 18.28\,$GeV$^2$ and $Q_0^2 =0.35\,$GeV$^2$. The value of 
$\Lambda^2$ characterizes the onset of the leading logarithmic
behavior from perturbative QCD. We note that the 
value for $\Lambda^2$ chosen here is about a factor of two
larger than in the earlier fits \cite{MMD,HMD}. 
This change is due to the strong deviation of the new
Jefferson Lab data for $G_E^p /G_M^p$ \cite{Jones:1999rz,Gayou:2001qd}
from the asymptotic prediction of perturbative QCD.

 The errors in the fit parameters quoted above are unknown.
 Due to the highly nonlinear nature of the problem (remember 
 that we fit all 4 form factors simultaneously) a trustworthy error 
 analysis is a very non-trivial problem. This is complicated by the fact 
 that the errors of some inputs (e.g. the two-pion continuum 
 \cite{Hohler:1974ht}) are not known. For this reason, no errors
 were given in the previous dispersion analyses by H\"ohler et al.
 \cite{Hohler:1976ax} and Mergell et al. \cite{MMD}
 (and also not in the widely used dispersion theoretical analysis of
 pion-nucleon scattering by H\"ohler and coworkers \cite{LB}). 
 At present, it is an open problem how to assign a serious theoretical error 
 to such type of analysis.

\begin{figure*}[htb]
\centerline{\includegraphics*[width=16cm,angle=0]{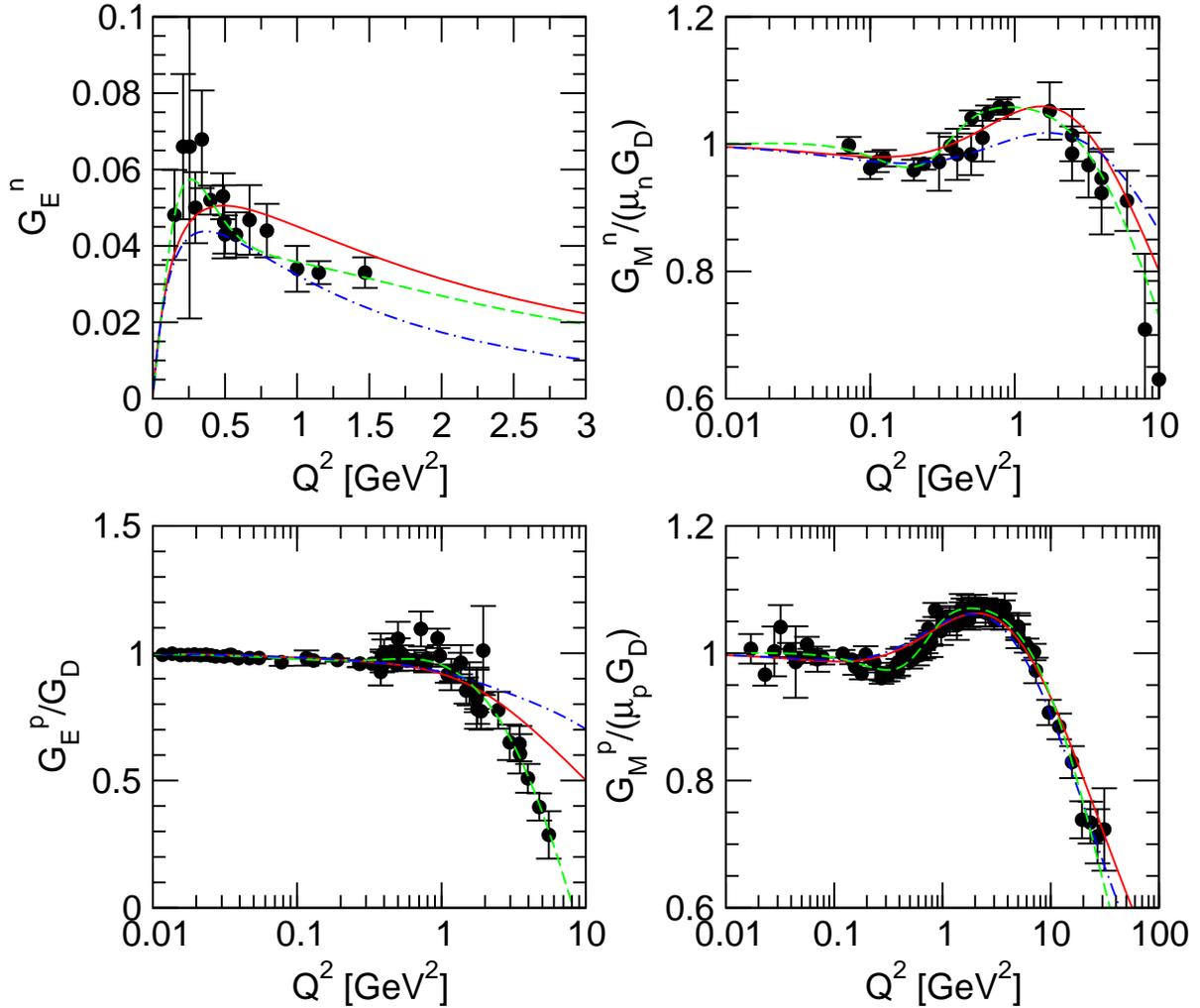}}
\caption[fig1]{\label{fig1}
The nucleon form factors for space-like momentum transfer. 
The solid line gives our best fit for  
$b_{ne} = (-1.33 \pm 0.027 \pm 0.03) \cdot 10^{-3}$~fm \protect\cite{Kop}
while the dash-dotted line gives fit 1 from Ref.~\protect\cite{HMD}. 
The dashed lines give the result of the phenomenological fit of 
Ref.~\protect\cite{FW}. 
}
\end{figure*}

In the previous dispersion analyses \cite{MMD,HMD}, 
a strong fine tuning of the residues of the excited $\rho$ mesons in 
the isovector channel could be observed. 
It is well known that the experimental dipole behavior of the 
form factors requires two narrow structures with opposite signs close to
each other. The fits in the previous analyses, however, tended to favor 
very large residues ($\sim 40$) for a small overall gain in $\chi^2$. 
Taken at face value, these residues implied unphysically large couplings for 
the isovector poles to the nucleon. In order to avoid this problem, 
we have restricted the magnitude of the vector meson coupling constants in 
the present analysis. As a consequence, the residues of all 
poles in the isoscalar and isovector channels are of order one.

In Fig.~\ref{fig1}, we show the resulting Sachs form factors by the solid 
line. The form factors are normalized to the
canonical dipole fit, with the exception of the neutron electric form factor,
which is not scaled. We can describe most of the data fairly well, with the
exception of the Jefferson Lab data on $G_E^p$ in the $Q^2$ range from 3 to 6
GeV$^2$ (see also the discussion in Refs.~\cite{Ulfff,Hammer:2001rh}). 
The $\chi^2$/datum of this fit is 2.07. For comparison, we exhibit
fit~1 of Ref.~\cite{HMD} by the dash-dotted line. This fit
is based on the 1996 data basis and describes the data clearly worse than our 
new fit. 
Also shown by the dashed line is the fit with the ``phenomenological 
ansatz'' of \cite{FW}, which has 6 free parameters for each of the 4 form 
factors.\footnote{We have reproduced this fit from 
Eqs.~(4-6) and the numbers given in Table 2 of the published version of 
Ref.~\cite{FW}. For $G_M^p$, we have used the 
following values for the parameters: $a_{10} = 1.0024$, $a_{11} = 0.7554$,
$a_{20} =  1.0000 - a_{10} = - 0.0024$, $a_{21} = 14.68$, $a_b = -0.159$,
$Q_b =  0.325$, $\sigma_b =  0.231$ which differ slightly from what is
given in Table 2 of the published version of Ref.~\cite{FW}. With the
numbers in that table, one can not reproduce the curve shown in Fig.~4 of
\cite{FW}. 
} 
It does significantly better for $G_E^p$. 
In Ref.~\cite{FW}, a number of other fits with a ``physically
motivated ansatz'' were also performed.
These fits are generally comparable in quality to the one shown in
Fig.~\ref{fig1} but give a better description of $G_M^p$. 

In the fits of Ref.~\cite{FW},
a pronounced structure around $Q^2 \simeq 0.2\,$GeV$^2$ was found 
for $G_E^n$ (and could also be isolated in the other form factors
as a small effect). 
\begin{figure}[htb]
\centerline{\includegraphics*[width=8cm,angle=0]{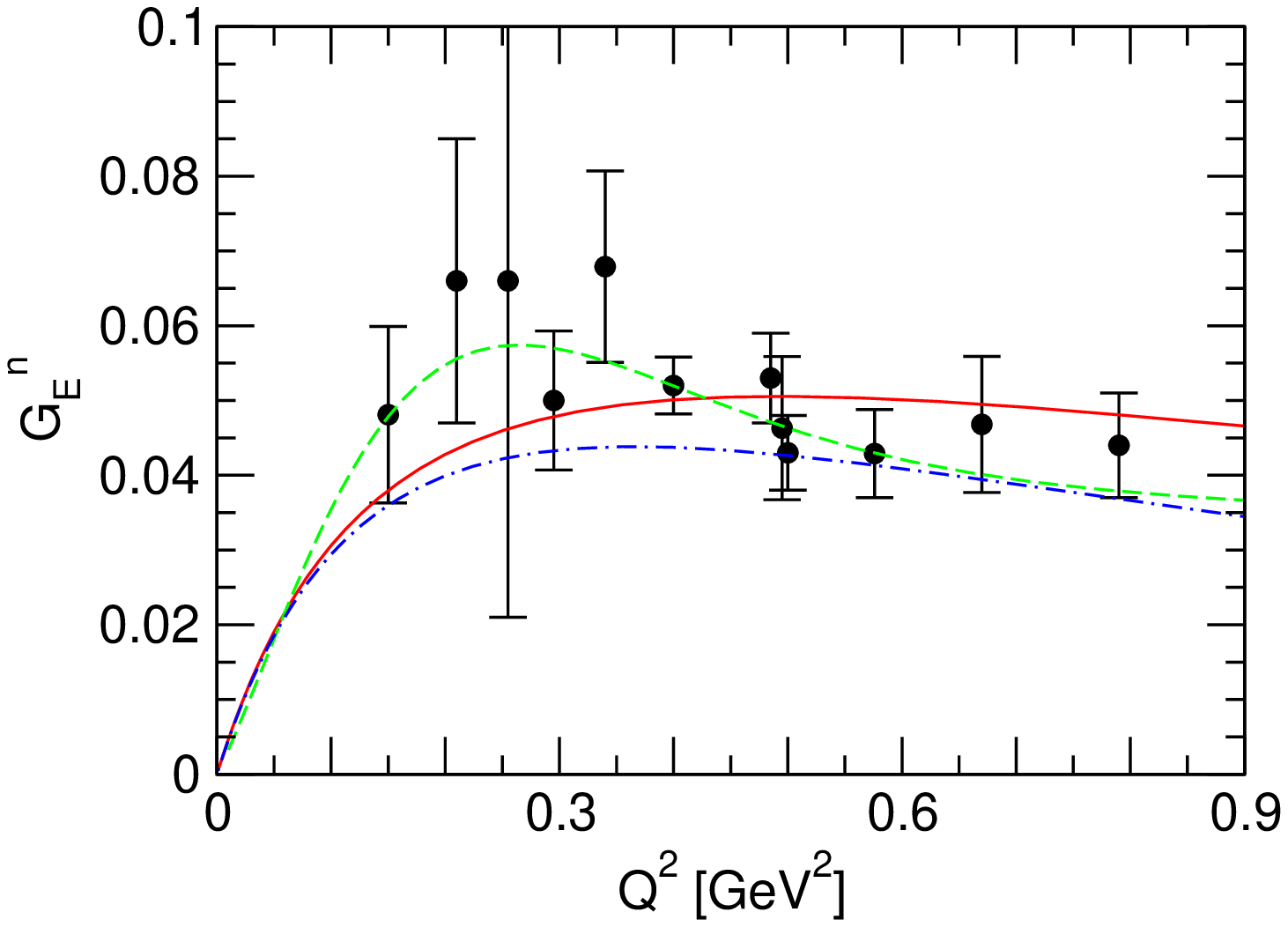}}
\caption[fig1]{\label{fig2}
The neutron charge form factor $G_E^n$ for momentum transfers
$Q^2=0...0.9$ GeV$^2$. The curves are as in Fig.~\ref{fig1}.
}
\end{figure}
This region of momentum transfers is shown in Fig.~\ref{fig2} in
greater detail. The structure was attributed to a contribution from
a very long range pion cloud (for a discussion of this point,
see also Ref.~\cite{HMD2}). Our analysis does not show the pronounced 
structure around  $Q^2 \simeq 0.2\,$GeV$^2$. 
What this result means remains to be seen. Taking the errors of the 
experimental data into account, the structure is not an unambiguous 
feature of the data (cf. Fig.~\ref{fig2}).
Ultimately, the question of whether the structure found in Ref.~\cite{FW}
is real physics or not should be decided by more accurate data 
in the region $Q^2 \simeq 0.2\,$GeV$^2$.
A similar plot for the proton charge form factor in the low momentum 
transfer region is shown in Fig.~\ref{fig3}. The data in this range
are sufficiently spread to leave room for various interpretations.
 \begin{figure}[htb]
\centerline{\includegraphics*[width=8cm,angle=0]{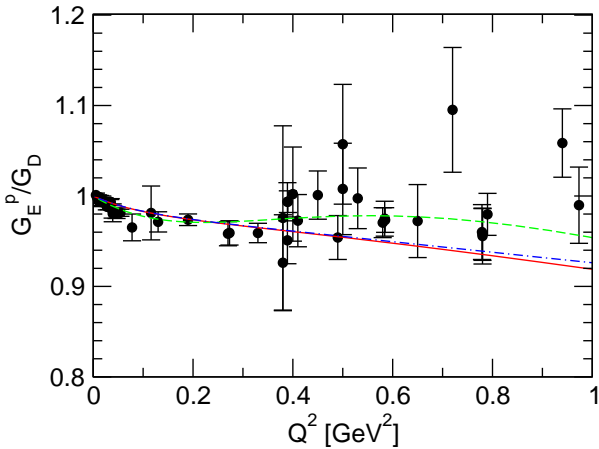}}
\caption[fig3]{\label{fig3}
The proton charge form factor $G_E^p$ for momentum transfers
$Q^2=0...1$ GeV$^2$. The curves are as in Fig.~\ref{fig1}.
}
\end{figure}

The $\omega NN$ and $\phi NN$ coupling constants derived from the 
residua given above are $g_1^{\omega NN}= 21.4$, $g_2^{\omega NN}= 0.9$,
$g_1^{\phi NN}= -10.3$, and $g_2^{\phi NN}= -3.1$. The absolute
values of these coupling constants are comparable with the results
of Refs.~\cite{MMD,HMD} but the small tensor couplings
$g_2^{\omega NN}$ and $g_2^{\phi NN}$
have changed sign. This change of sign indicates that the error in 
the extracted vector meson coupling constants is of the order of the 
tensor couplings.

For the electric and magnetic radii of the nucleon, we find the 
values $r_E^p = 0.848$ fm, $r_M^p = 0.857$ fm, and $r_M^n = 0.879$ fm.
As in the previous analysis \cite{MMD,HMD}, the proton charge radius is 
somewhat small compared to the recent precise determinations
$r_E^p = 0.880(15)$ fm \cite{Rosenfelder} and $r_E^p = 0.895(18)$ fm
\cite{Sick:2003gm} from low-momentum-transfer data and
$r_E^p = 0.883(14)$ fm \cite{Melnikov:1999xp} from Lamb shift
measurements.

Note that the recent determinations of the proton charge radius 
by Rosenfelder \cite{Rosenfelder} and Sick \cite{Sick:2003gm} are
based on analyzing low-momentum cross section data directly and
include corrections from two-photon exchange. Our analysis relies on
form factor data extracted in the one-photon exchange approximation
where the two-photon corrections enter in the systematical error.
A comparison of our value to the analyses of low-momentum transfer cross 
sections directly allows one to estimate systematic errors in the extraction 
of the radii. Of course, one might contemplate a fit to cross section 
data only. However, for consistency
the two-photon corrections have to be applied to the full 
data basis (as opposed to low $Q^2$ data only) and this will certainly take a 
long time. This issue has to be addressed in the future but requires a large 
common effort by experimenters and theorists.

Our results for the magnetic radii of the proton and neutron are consistent 
with the recent values from continued fraction expansions:
Kubon et al.~\cite{Kubon:2001rj} extracted $r_M^n = 0.873(11)$ fm 
from precise data for $G_M^n$, while Sick \cite{Sick:2003gm,Sickprivate}
obtained $r_M^p = 0.855(35)$ fm from a careful analysis of the world
data on elastic electron-proton scattering. To simplify the comparison,
we have collected our results and the other recent radius determinations 
in Table \ref{tab_rad}.
\renewcommand{\arraystretch}{1.2}
\begin{table}[htb]
\begin{center}
\begin{tabular}{|c|c|c|c|}
\hline\hline
$r_E^p$ [fm] &  $r_M^p$ [fm] & $r_M^n$ [fm] & Reference \\ \hline 
0.848 &  0.857 & 0.879 & this work\\
0.880(15) & & & \protect\cite{Rosenfelder}\\
0.895(18) & 0.855(35) & & \protect\cite{Sick:2003gm,Sickprivate}\\
0.883(14) & & &  \protect\cite{Melnikov:1999xp}\\
 & & 0.873(11) & \protect\cite{Kubon:2001rj} \\ \hline\hline
\end{tabular}
\vspace{0.3cm}
\caption{\label{tab_rad}
Comparison of our results for the radii $r_E^p$, $r_M^p$, and $r_M^n$ 
with other recent determinations. The numbers in parentheses
indicate the error in the last digits.
}
\end{center}
\end{table}

\section{Outlook}
\label{sec:out}

We have shown that a dis\-per\-sion-theo\-reti\-cal analysis 
based on a minimal number of poles can describe most of the 
current world data on the nucleon electromagnetic
form factors for space-like momentum transfer.
While the charge radius of the proton is still somewhat small,
the magnetic radii of the proton and neutron are in good agreement with the 
recent determinations using continued fraction expansions
\cite{Sick:2003gm,Kubon:2001rj,Sickprivate}.
 
The spectral functions used here can be improved. Ideally,
one would like spectral functions consisting of various continuum
contributions with vector mesons emerging naturally as resonances
and avoid explicit pole terms altogether.
In particular, the isoscalar region around masses of 1~GeV could be 
supplemented by explicit $\pi\rho$ \cite{Meissner:1997qt} and $K\bar K$ 
\cite{HaRM} continuum contributions.
The $\phi$ meson would then appear as a finite 
width resonance in the $K\bar K$ (and possibly $\pi\rho$) continua,
thereby eliminating the need to include it explicitly as a pole term.
It would also be desirable to replace the poles with masses in the 
region from 1.5 to 2.0 GeV with an appropriate continuum contribution.
However, the analytical continuation of experimental
data that is required to obtain the continua becomes more and more 
difficult the higher one gets in mass, so that the latter goal is not 
realistic in the near future.
Another improvement is to construct a better representation of the 
perturbative QCD behaviour  for large $Q^2$. 
This would allow to include all existing time-like data in the 
fits, and thus lead to a more consistent description of these fundamental 
quantities.

Furthermore, the proton charge radius extracted from 
low-momentum-transfer data and Lamb shift measurements could be included
as a further constraint in the analysis.
This would allow to check the consistency between the high-$Q^2$
data and the low-$Q^2$ extractions of the radii.
Whether this requires the introduction of additional vector meson poles is 
an open question. Last but not least, a full error analysis of the 
extracted radii and coupling constants should be carried out.
As mentioned above,
this is a nontrivial task because of the highly nonlinear nature of the 
problem. Work along these lines is under way \cite{BHM}.

\section*{Acknowledgment}
We acknowledge valuable discussions with D. Drechsel, J. Fried\-rich, 
I. Sick, and Th. Walcher. Furthermore, we thank 
J. Fried\-rich and Th. Walcher for providing us with their data basis
and updated fit parameters.

\end{document}